%% file: main.tex
\documentclass[sigconf]{acmart}

\usepackage[caption=false]{subfig}

\AtBeginDocument{%
  \providecommand\BibTeX{{%
    \normalfont B\kern-0.5em{\scshape i\kern-0.25em b}\kern-0.8em\TeX}}}

\setcopyright{acmcopyright}
\acmPrice{15.00}
\acmDOI{10.1145/3445814.3446763}
\acmYear{2021}
\copyrightyear{2021}
\acmSubmissionID{asplos21main-p1455-p}
\acmISBN{978-1-4503-8317-2/21/04}
\acmConference[ASPLOS '21]{Proceedings of the 26th ACM International Conference on Architectural Support for Programming Languages and Operating Systems}{April 19--23, 2021}{Virtual, USA}
\acmBooktitle{Proceedings of the 26th ACM International Conference on Architectural Support for Programming Languages and Operating Systems (ASPLOS '21), April 19--23, 2021, Virtual, USA}

\begin{document}

\newcommand{\Name}{RecSSD}
\title[\Name]{\Name:  Near Data Processing for \\ Solid State Drive Based Recommendation Inference}


\author{Mark Wilkening}
\affiliation{%
  \institution{Harvard University}
  \streetaddress{33 Oxford St.}
  \city{Cambridge}
  \state{Massachusetts}
  \country{USA}}
\email{wilkening@g.harvard.edu}

\author{Udit Gupta}
\affiliation{%
  \institution{Harvard University \\ Facebook}
  \streetaddress{33 Oxford St.}
  \city{Cambridge}
  \state{Massachusetts}
  \country{USA}}
\email{ugupta@g.harvard.edu}

\author{Samuel Hsia}
\affiliation{%
  \institution{Harvard University}
  \streetaddress{33 Oxford St.}
  \city{Cambridge}
  \state{Massachusetts}
  \country{USA}}
\email{shsia@g.harvard.edu}

\author{Caroline Trippel}
\affiliation{%
  \institution{Facebook}
  \streetaddress{1 Hacker Way}
  \city{Menlo Park}
  \state{California}
  \country{USA}}
\email{ctrippel@fb.com}

\author{Carole-Jean Wu}
\affiliation{%
  \institution{Facebook}
  \streetaddress{1 Hacker Way}
  \city{Menlo Park}
  \state{California}
  \country{USA}}
\email{carolejeanwu@fb.com}

\author{David Brooks}
\affiliation{%
  \institution{Harvard University}
  \streetaddress{33 Oxford St.}
  \city{Cambridge}
  \state{Massachusetts}
  \country{USA}}
\email{dbrooks@eecs.harvard.edu}

\author{Gu-Yeon Wei}
\affiliation{%
  \institution{Harvard University}
  \streetaddress{33 Oxford St.}
  \city{Cambridge}
  \state{Massachusetts}
  \country{USA}}
\email{guyeon@seas.harvard.edu}

\renewcommand{\shortauthors}{Wilkening, et al.}

\begin{abstract}
Neural personalized recommendation models are used across a wide variety of datacenter applications including search, social media, and entertainment.
State-of-the-art models comprise large embedding tables that have billions of parameters requiring large memory capacities.
Unfortunately, large and fast DRAM-based memories levy high infrastructure costs. 
Conventional SSD-based storage solutions offer an order of magnitude larger capacity, but have worse read latency and bandwidth, degrading inference performance.
\Name\ is a near data processing based SSD memory system customized for neural recommendation inference that reduces end-to-end model inference latency by 2$\times$ compared to using COTS SSDs across eight industry-representative models.
\end{abstract}

\ccsdesc[500]{Hardware~External storage}
\ccsdesc[500]{Computer systems organization~Neural networks}

\keywords{near data processing, neural networks, solid state drives}


\maketitle

\input{introduction}

\input{background}

\input{characterization}
\input{design}
\input{methodology}
\input{evaluation}

\input{related}
\input{conclusion}

\begin{acks}
We would like to thank the anonymous reviewers for their thoughtful comments and suggestions.
We would also like to thank Glenn Holloway for his
valuable technical support.
This work was sponsored in part by National Science Foundation Graduate Research Fellowships (NSFGRFP), and the ADA (Applications Driving Architectures) Center.
\end{acks}

\appendix

\input{artifact}

\bibliographystyle{ACM-Reference-Format}
\bibliography{refs}

\end{document}

%% file: introduction.tex
\section{Introduction}
Recommendation algorithms are used across a variety of Internet services such as social media, entertainment, e-commerce, and search~\cite{gupta2019architectural, mtwnd, dinzhou2018deep,dienzhou2019deep, mckinsey}.
In order to efficiently provide accurate, personalized, and scalable recommendations to users, state-of-the-art algorithms use deep learning based solutions.
These algorithms consume a significant portion of infrastructure capacity and cycles in industry datacenters.
For instance, compared to other AI-driven applications, recommendation accounts for 10$\times$ the infrastructure capacity in Facebook's datacenter~\cite{gupta2019architectural,naumov2019dlrm,lui2020understanding}.
Similar capacity requirements can be found at Google, Alibaba, and Amazon~\cite{mtwnd, dinzhou2018deep, dienzhou2019deep}.


One of the key distinguishing features of neural recommendation models is processing categorical input features using large embedding tables.
While large embedding tables enable higher personalization, they consume up to hundreds of GBs of storage~\cite{gupta2019architectural,park2018deep}.
In fact, in many cases, the size of recommendation models is set by the amount of memory available on servers~\cite{gupta2019architectural}. 
A promising alternative is to store embedding tables in SSDs.
While SSDs offer orders of magnitude higher storage capacities than main memory systems, they exhibit slower read and write performance.
To hide the longer SSD read and write latencies, previous SSD based systems overlap computations from other layers in the recommendation models and cache frequently accessed embedding vectors in DRAM-based main memory~\cite{eisenman2018bandana,Zhao2020DistributedHG, Zhao:CIKM}. 




We propose {\it \Name}, a near data processing (NDP) solution customized for recommendation inference that improves the performance of the underlying SSD storage for embedding table operations. 
In order to fully utilize the internal SSD bandwidth and reduce round-trip data communication overheads between the host CPU and SSD memory, \Name\ offloads the entire embedding table operation, including gather and aggregation computations, to the SSDs. 
Compared to baseline SSD, we demonstrate that \Name\ provides a 4$\times$ improvement in embedding operation latency and 2$\times$ improvement in end-to-end model latency on a real OpenSSD system.
In addition to offloading embedding operations, \Name\ exploits the locality patterns of recommendation inference queries. 
\Name\ demonstrates that a combination of host-side and SSD-side caching complement NDP and reduce end-to-end model inference latency. 
To demonstrate the feasibility and practicality of the proposed design in server-class datacenter systems, We implement \Name~ on a real, open-source Cosmos+OpenSSD system within the Micron UNVMe driver library.

The key contributions of this paper are: 

\begin{itemize}


  
    \item We design \Name, the first NDP-based SSD system for recommendation inference.
    Improving the performance of conventional SSD systems, the proposed design targets the main performance bottleneck to datacenter scale recommendation execution using SSDs. 
    Furthermore, the latency improvement further enables recommendation models with higher storage capacities at reduced infrastructure cost.
    
  \item We implement \Name~in a real system on top of the Cosmos+OpenSSD hardware. 
    The implementation demonstrates the viability of Flash-based SSDs for industry-scale recommendation. In order to provide a feasible solution for datacenter scale deployment, we implement RecSSD within the FTL firmware; the interface is compatible with existing NVMe protocols, requiring no hardware changes.


  \item We evaluate the proposed design across eight industry representative models across various use cases (e.g., social media, e-commerce, entertainment).
    Of the eight, our real system evaluation shows that five models --- whose runtime is dominated by compute-intensive FC layers --- achieve comparable performance using SSD compared to DRAM.
    The remaining three models are dominated by memory-bound, embedding table operations.
    On top of the highly optimized hybrid DRAM-SSD systems, we demonstrate that \Name~improves performance by up to 4$\times$ for individual embedding operations, translating into up to 2$\times$ end-to-end recommendation inference latency reduction.

  
\end{itemize}

%% file: background.tex
\section{Background}~\label{sec:back}
\subsection{Recommendation Systems} \label{sec:back:recsys}
Often found in commercial applications, recommendation systems recommend \textit{items} to \textit{users} by predicting said items' \textit{values} in the context of the users' \textit{preferences}. In fact, meticulously tuned personalized recommendation systems form the backbone of many internet services -- including social media, e-commerce, and online entertainment~\cite{dinzhou2018deep, dienzhou2019deep, naumov2019dlrm, ncf, mtwnd} -- that require real-time responses.
Modern recommendation systems implement deep learning-based solutions that enable more sophisticated user-modeling. 
Recent work shows that deep-learning based recommendation systems not only drive product success\cite{mckinsey, chui2018notes, microsoftPersonalizedRec} but also dominate the datacenter capacity for AI training and inference~\cite{gupta2019architectural, kim2018hpca,Naumov2020DeepLT}.
Thus, there exists a need to make dataceter-scale recommendation solutions more efficient and scalable.
\begin{figure}[ht]
    \centering
    \includegraphics[width=\linewidth]{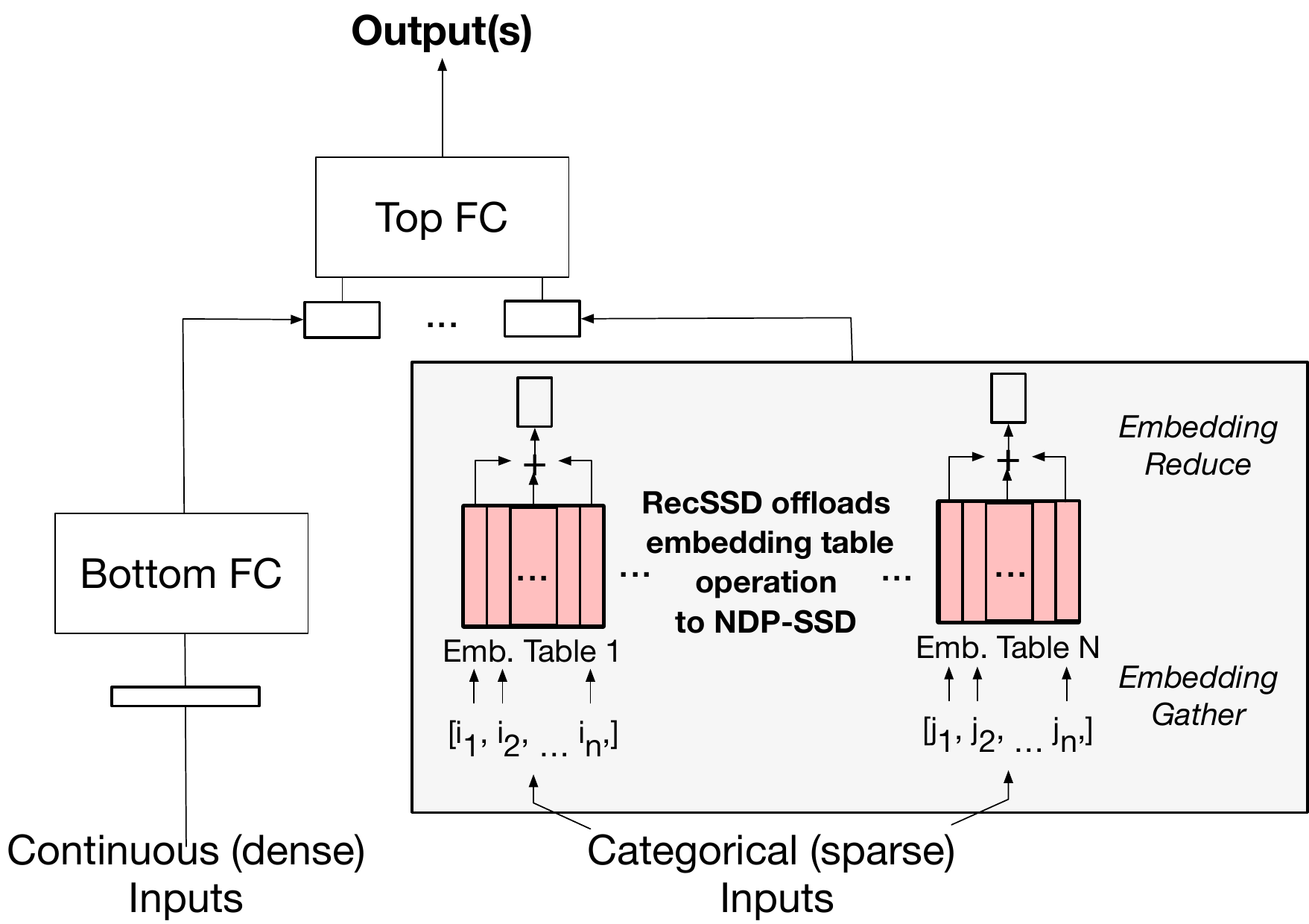}
    \caption{Recommendation models process both categorical and continuous input features. 
    }
    \label{fig:rec}
\end{figure}

\textbf{Overview of model architecture} As shown in Figure~\ref{fig:rec}, deep learning-based recommendation models comprise both fully-connected (FC) layers and embedding tables of various sizes. 
FC layers stress compute capabilities by introducing regular MAC operations while embedding table references stress memory bandwidth by introducing irregular memory lookups.
Based on the specific operator composition and dimensions, recommendation models span a diverse range of architectures. 
For instance, the operators that combine outputs from Bottom FC and embedding table operations depend on the application use case. 
Furthermore, recommendation models implement a wide range of sizes of FC layers and embedding tables.
%

\textbf{Processing categorical inputs}
Unique to recommendations, models process categorical input features using embedding table operations. 
Embedding tables are organized such that each row is a unique embedding vector typically comprising 16, 32, or 64 learned features (i.e., number of columns for the table).
For each inference, a set of embedding vectors, specified by a list of IDs (e.g., multi-hot encoded categorical inputs) is \textit{gathered} and \textit{aggregated} together.
Common operations for aggregating embedding vectors together include sum, averaging, concatentation, and matrix multiplication~\cite{naumov2019dlrm,mtwnd,dinzhou2018deep,dienzhou2019deep}; Figure~\ref{fig:rec} shows an example using summation. 
Inference requests are often batched together to amortize control overheads and better utilize computational resources. 
Additionally models often comprise many embedding tables. 
Currently, production-scale datacenter store embedding tables in DRAM while CPU perform embedding table operations, optimizations such as vectorized instructions and software prefetching~\cite{Caffe2SLS}. 

The embedding table operations pose unique challenges:

\begin{enumerate}
    \item \textbf{Capacity:} Industry-scale embedding tables have up to hundreds of millions of rows leading to embedding tables that often require up to $\sim$10GBs of storage~\cite{gupta2019architectural}. In fact, publications from industry illustrate that the aggregate capacity of all embedidng tables in a neural recommendation model can require TBs of storage~\cite{Zhao2020DistributedHG,Zhao:CIKM}
    
    
    \item \textbf{Irregular Accesses:} Categorical input features are sparse, multi-hot encoded vectors. 
    High sparsity leads to a small fraction of embedding vectors being access per request. Furthermore, access patterns between subsequent requests from different users can be quite different causing embedding table operations to incur irregular accesses. 
    
    
    \item \textbf{Low Compute Intensity:} The overall compute intensity of the embedding tables are orders of magnitude lower than other deep learning workloads precluding efficient execution using recently proposed SIMD, systolic array, and dataflow hardware accelerators~\cite{gupta2019architectural}. 
\end{enumerate}

These three features -- large capacity requirements, irregular memory accesses, and low compute intensity -- make Flash technology an interesting target for embedding tables.


\subsection{Flash memory systems}
\label{sec:back:flash}

\begin{figure}[ht]
    \centering
    \includegraphics[width=\linewidth]{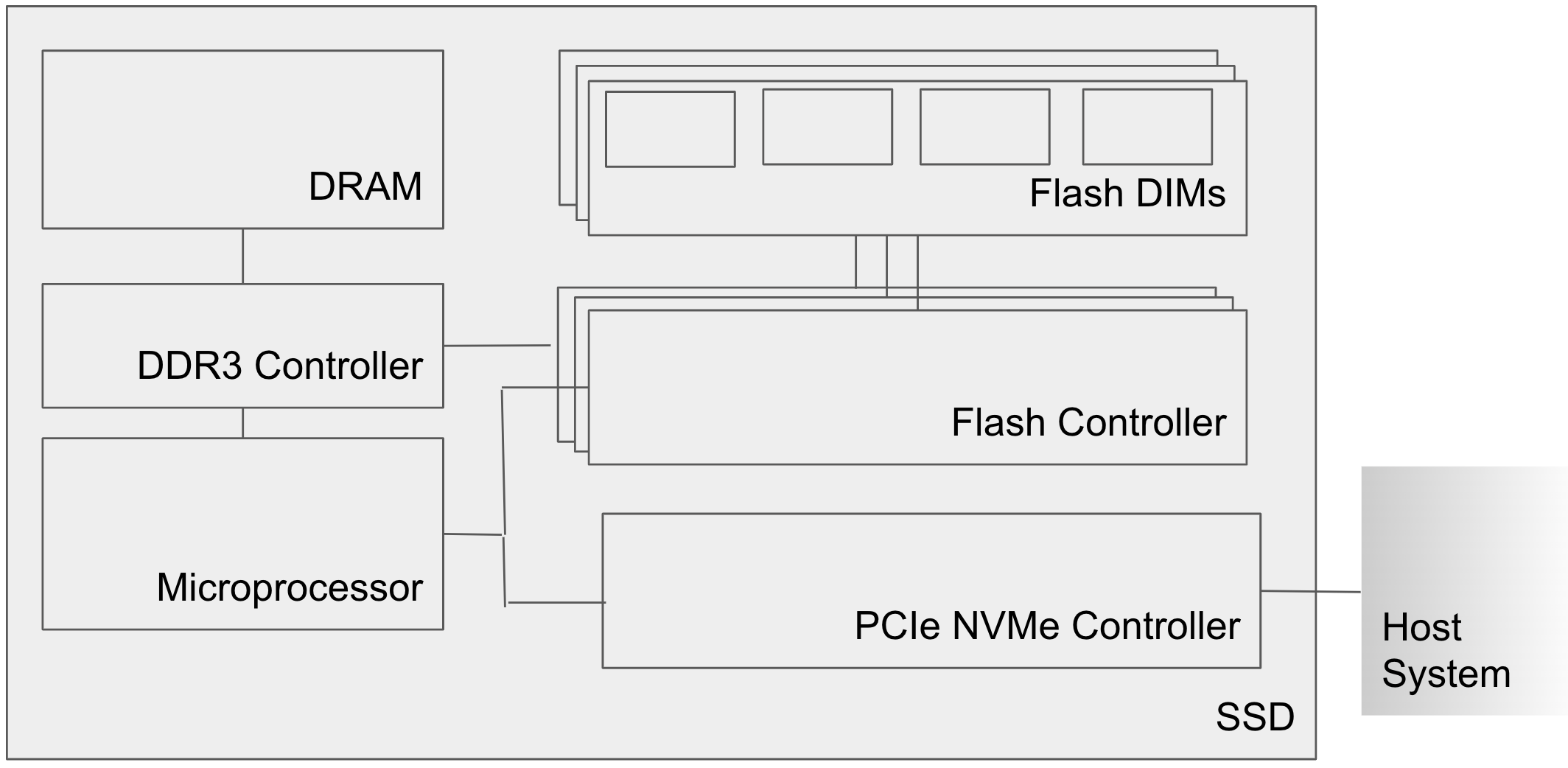}
    \caption{In order to support a high performance and simple logical block interface to the host while handling the peculiarities of NAND Flash memories, SSDs are designed with a Microprocessor operating alongside dedicated memory controllers. }
    \label{fig:ssd_architecture}
\end{figure}

\textbf{Architecture of Flash} 
NAND flash memory is the most widely used SSD building block on the market.
Compared to traditional disk-based storage systems, NAND flash memories offer higher performance in terms of latency and bandwidth for reads and writes~\cite{andersen2010rethinking}.
Figure~\ref{fig:ssd_architecture} illustrates the overall architecture of NAND Flash storage systems.
To perform a read operation, the host communicates over PCIe using an NVMe protocol to a host controller on the SSD. 
The host requests logical blocks, which are served by a \textit{flash translation layer} (FTL) running on a microprocessor on the SSD. 
The FTL schedules and controls an array of Flash controllers, which are organized per channel and provide specific commands to all the Flash DIMs (chips) on a channel and DMA capabilities across the multiple channels. 
In order to transfer data between the Flash controller's DMA engines and the host NVMe DMA engine, the controller uses an on-board DRAM buffer. 

\textbf{Flash Translation Layer (FTL)}
In order to maintain compatibility with existing drivers and file systems, Flash SSD systems implement the FTL. 
The FTL exposes a logical block device interface to the host system while managing the underlying NAND Flash memory system. 
This includes performing key functions such as (1) maintaining indirect mapping between logical and physical pages, (2) maintaining a log-like write mechanism to sequentially add data in erase blocks and invalidate stale data~\cite{Rosenblum1992}, (3) garbage collection, and (4) wear leveling.
As shown in Figure~\ref{fig:ssd_architecture}, to perform this diverse set of tasks, the FTL runs on a general purpose microprocessor.

\textbf{Performance characteristics of SSD storage} 
Compared to DRAM-based main memory systems, Flash-based storage systems have orders of magnitude higher storage densities~\cite{andersen2010rethinking} enabling higher capacities at lower infrastructure cost, around 4-8x cheaper than DRAM per bit~\cite{eisenman2018}.
Despite these advantages, Flash poses many performance challenges. 
One single flash memory package provides a limited bandwidth of 32-40MB/sec~\cite{Agrawal2008}. 
In addition, writes to flash memory are often much slower, incurring \textit{$O$(ms)} latencies.
To help address these limitations, SSDs are built to expose significant internal bandwidth by organizing flash memory packages as an array of connected channels (e.g., 2-10) handled by a single memory controller. 
Since logical blocks can be striped over multiple flash memory packages, data accesses can be conducted in parallel to provide higher aggregated bandwidth and hide high latency operations through concurrent work.


%




%% file: characterization.tex
\section{
SSD Storage for Neural Recommendation}
~\label{sec:characterize}
To better understand the role of SSD storage for neural recommendation inference, we begin with initial characterization. 
First, we present the memory access pattern characterization for recommendation models running in a cloud-scale production environment
and describe the locality optimization opportunities for performing embedding execution on SSDs. 
Then, we take a step further to study the impact of storing embedding tables and performing associated computation in SSDs as opposed to DRAM~\cite{Ke2019RecNMPAP}.
The characterization studies focus on embedding table operations, followed by the evaluation on the end-to-end model performance. 

\subsection{Embedding access patterns in production models}~\label{sec:characterize:locality}
One important property of SSD systems is that SSDs operate as block devices where data is transferred in coarse chunks. This is an important factor when considering efficient bandwidth use of SSDs. The hardware is designed for sequential disk access, where data is streamed in arbitrarily large chunks. However, larger access granularity penalizes performance for workloads that require random, sparse accesses -- embedding table access and operation in neural recommendation models. 
Therefore, it is important to understand unique memory access patterns of embedding tables. Furthermore, caching techniques become even more important to exploit temporal reuse and maximize spatial locality from block accesses. 

\begin{figure}[t]
    \centering
    \includegraphics[width=\linewidth]{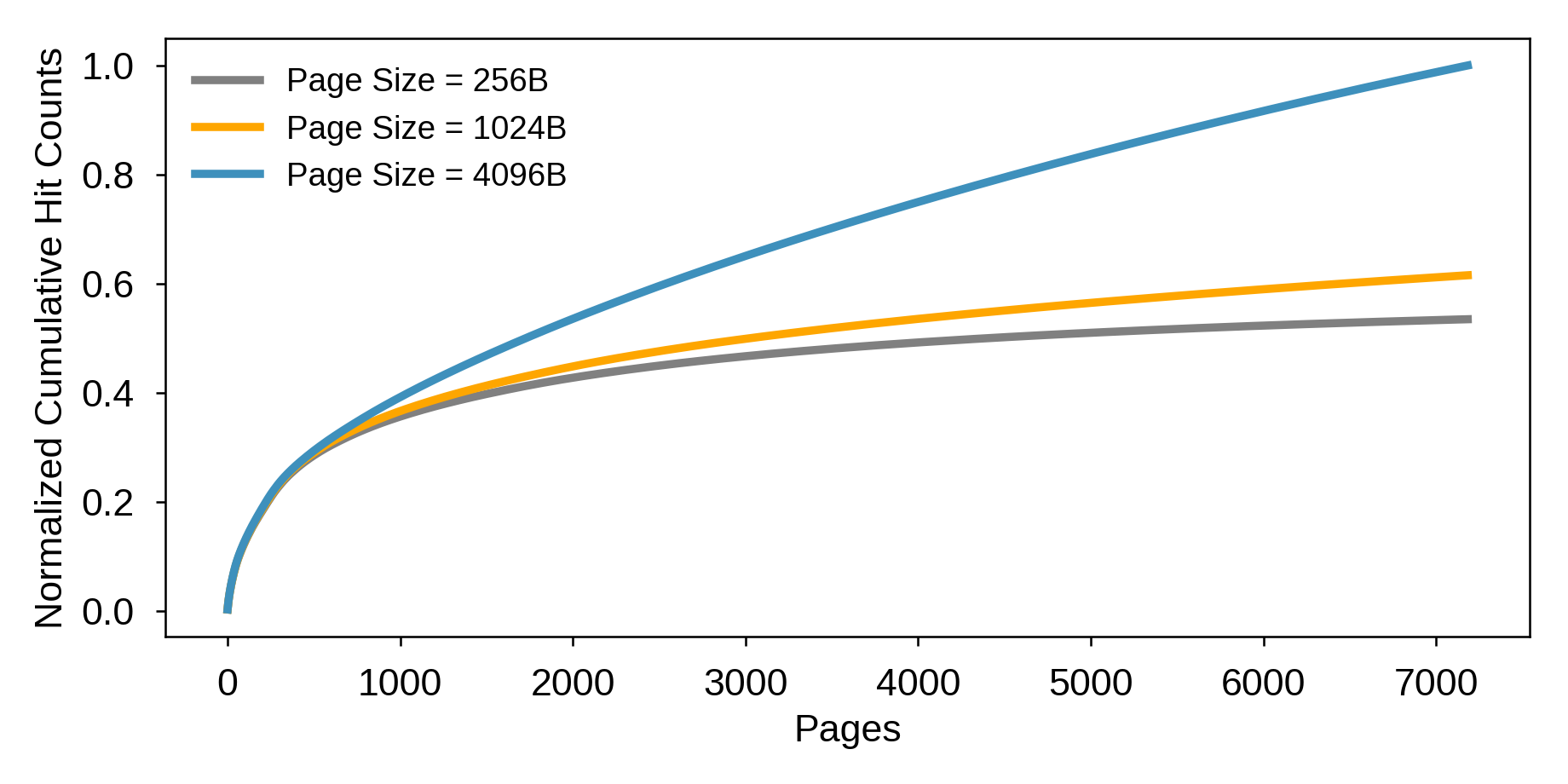}
    \caption{Access patterns to neural recommendation embedding tables follow the power-law distribution.}
    \label{fig:embedding_example0}
\end{figure}

Figure~\ref{fig:embedding_example0} depicts the reuse distribution of embedding tables in the granularity of 256B, 1KB, and 4KB, respectively. The x-axis represents pages accessed over the execution of real-time recommendation inference serving (sorted by the corresponding hit counts in the ascending order) whereas the y-axis shows the cumulative hit counts, by analyzing embedding table accesses logged for recommendation models running in a cloud-scale production environment.
Access patterns to embedding tables follow the power-law distribution.
Depending on the page sizes, the slope of the tail changes. 
\textit{The majority of reuse remains concentrated in a few hot memory regions --- a few hundred pages capture 30\% of reuses while caching a few thousand pages can extend reuse over 50\%.}

\begin{figure}[t]
    \centering
    \includegraphics[width=\linewidth]{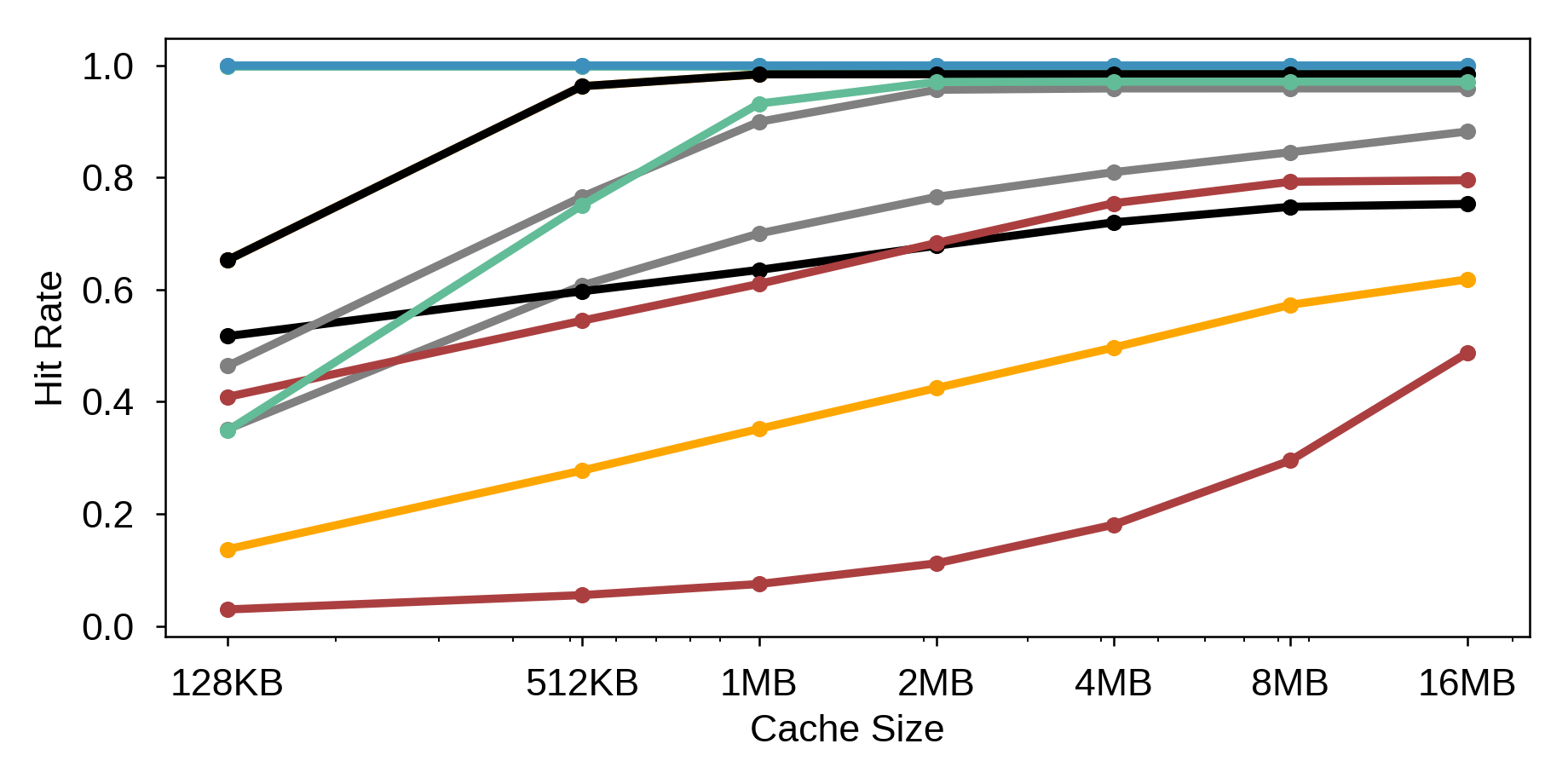}
    \caption{Locality patterns vary significantly across different embedding tables. Using a 16-way LRU 4KB page cache of varying total capacities, the hit rate varies wildly from under 10\% to over 90\% across different embedding tables.
    }
    \label{fig:embedding_locality}
\end{figure}

The concentration of hot pages varies across individual embedding tables. Figure~\ref{fig:embedding_locality} characterizes the memory locality patterns across different, individual embedding tables. 
Using a 16-way, LRU, 4KB page cache of varying cache capacities, the hit rate varies wildly from under 10\% to over 90\% across the different embedding tables of recommendation models running in a cloud-scale production environment. 
As the capacity of the page cache increases, more embedding vectors can be captured in the page cache, leading to higher reuses. With a 16MB page cache per embedding table, more than 50\% of reuses can be achieved across all the embedding tables analyzed in this study. The specific page cache capacity per embedding table can be further optimized for better efficiency.   

Locality in embedding table accesses influences the design and performance of SSD systems in many ways.
First, on-board SSD caching is difficult due to the limited DRAM capacity and the potentially large reuse distances. 
Despite this, the distribution of reuse and the relatively small collection of hot pages suggest reasonable effectiveness of static partitioning strategies, where hot pages can be stored in host-side DRAM. 
But, most importantly, the varying page reuse patterns (Figure~\ref{fig:embedding_locality}) suggests that, although in some cases, caching can be used to effectively deal with block access, strategies for more efficiently handling sparse access is also needed. Previous work~\cite{eisenman2018bandana} has thoroughly investigated advanced caching techniques, while we propose orthogonal solutions which specifically target increasing the efficiency of sparse accesses. 
We evaluate our proposed techniques by using somewhat simpler caching strategies (standard LRU software caching and static partitioning) and sweeping the design space across a variety of input locality distributions.

\begin{figure}[t]
    \centering
    \includegraphics[width=\linewidth]{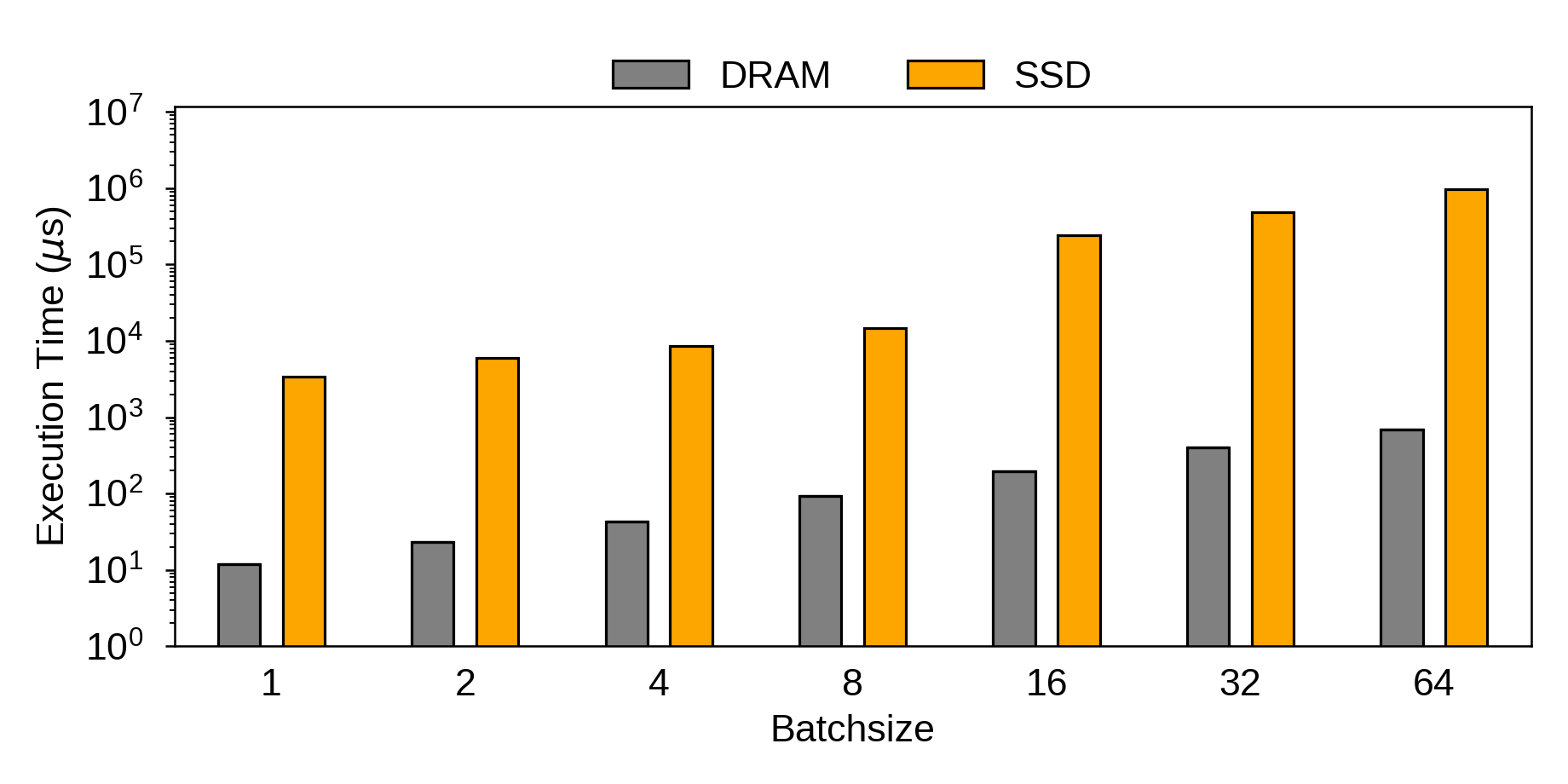}
    \caption{
    Using a table configuration typical of industry scale models~\cite{Gupta2020DeepRecSysAS, gupta2019architectural}
    and a range of batch sizes, the Sparse Length Sum (SLS) embedding table operation slows down significantly using SSD storage over DRAM.}
    \label{fig:baseline_characterization}
\end{figure}

\subsection{Performance of individual embedding operations }
\label{sec:baseline_characterization}

Given their unique memory access patterns, storing embedding tables in SSD versus DRAM has a large impact on performance given the characteristics of the underlying memory systems.
Figure~\ref{fig:baseline_characterization} illustrates the performance of a single embedding table operation using DRAM versus SSD across a range of batch-sizes.
The embedding table has one million rows, with an embedding vector dimension of 32, and 80 lookups per table, typical for industry-scale models such as Facebook's embedding-dominated recommendation networks~\cite{Gupta2020DeepRecSysAS, gupta2019architectural}.
For an optimized DRAM-based embedding table operation, we analyze the performance of the SparseLengthsSum operation in Caffe2~\cite{Caffe2}. 
As shown in Figure~\ref{fig:baseline_characterization}, compared to the DRAM baseline, accessing embedding tables stored in the SSD incurs three orders of magnitude longer latencies.
This is a result of software overhead in accessing embedding tables over PCIe as well as the orders-of-magnitude lower read bandwidth in the underlying SSD system --- 10K IOPS or 10$MB/s$ random read bandwidth on SSD versus 1$GB/s$ on DRAM. 
Thus, while SSD storage offers appealing capacity advantage for growing industry neural recommendation models, there is significant room to improve the performance of embedding table operations.


\subsection{Performance of end-to-end recommendation models}
While embedding tables enable recommendation systems to more accurately model user interests, as shown in Figure~\ref{fig:rec}, embedding is only a component when considering end-to-end recommendation inference.
Thus, to understand the end-to-end performance impact by offloading embedding tables to the SSD memory, we characterize the performance impact on recommendation inference over a representative variety of network model architectures. 


Our evaluations use eight open-source recommendation models~\cite{Gupta2020DeepRecSysAS} representing industry-scale inference use cases from Facebook, Google, and Alibaba~\cite{dinzhou2018deep, dienzhou2019deep, mtwnd, naumov2019dlrm, ncf, gupta2019architectural}.
For the purposes of this study, models are clustered into two categories based on the respective performance characteristics: embedding-dominated and MLP-dominated.
MLP-dominated models, such as Wide and Deep (WD), Multi-Task Wide and Deep (MTWND), Deep Interest (DIN), Deep Interest Evolution (DIEN), and Neural Collaborative Filtering (NCF), spend the vast majority of their execution time on matrix operations.
On the other hand, embedding-dominated models, such as DLRM-RMC1, DLRM-RMC2, and DLRM-RMC3, spend the majority of their time processing embedding table operations. We refer the reader to~\cite{Gupta2020DeepRecSysAS} for detailed operator breakdowns and benchmark model characterizations.

\begin{figure}[t]
    \centering
    \includegraphics[width=\linewidth]{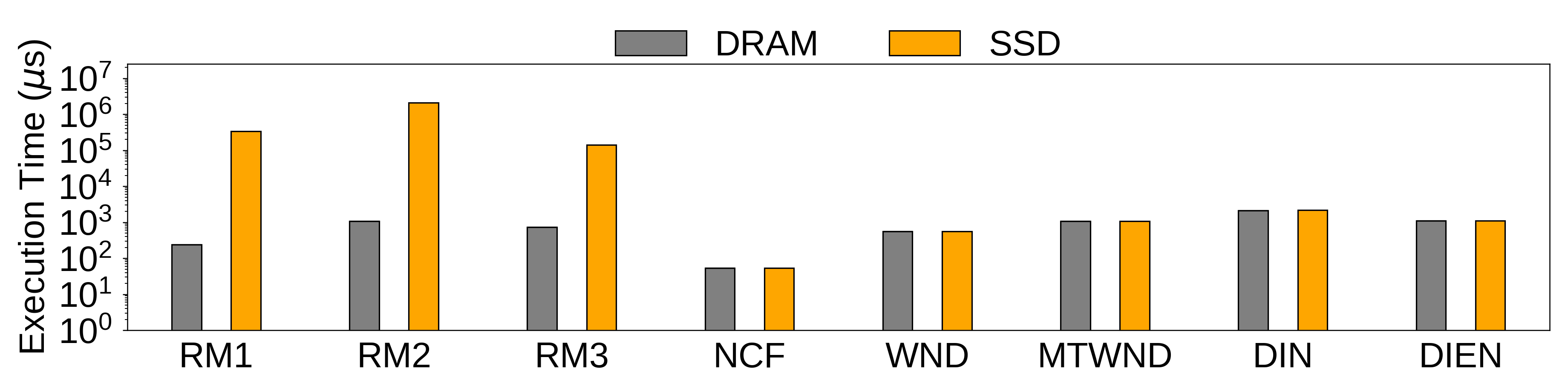}
    \caption{
    Performance degradation from using Flash based embedding table operations is model dependant. Storing tables in SSD for WND, MTWND, DIEN, and NCF increases model latency by 1.01$\times$, 1.01$\times$,  1.09$\times$, and 1.01$\times$, versus DRAM.}
    \label{fig:baseline_endtoend}
\end{figure}

Figure~\ref{fig:baseline_endtoend} shows the execution time of the eight recommendation models at a batch-size of 64 when embedding tables are stored in DRAM and in SSD, respectively.
The execution time for MLP-dominated models remains largely unaffected between the two memory systems.
Compared to DRAM, storing tables in SSD for WND, MTWND, DIEN, and NCF increases the model latency by 1.01$\times$, 1.01$\times$,  1.09$\times$, and 1.01$\times$.
On the other hand, storing embedding tables in SSD instead of DRAM significantly impacts the execution time for embedding-dominated models.
For instance, the execution time of embedding-dominated models, such as DLRM-RMC1, DLRM-RMC2, DLRM-RMC3, degrades by several orders of magnitude. 

\subsection{Opportunities for acceleration and optimization}
Given the performance characterization of individual embedding operations and end-to-end models when embeddings are stored in SSDs, we identify many opportunities for inference acceleration and optimization.
First, the overwhelming majority of execution time in MLP-dominated models is devoted to matrix operations; thus SSDs systems offer an exiting solution to store embedding tables in high-density storage substrates, lowering infrastructure costs for datacenter scale recommendation inference.

While SSDs is an appealing target for MLP-dominated models, there is significant room for performance improvement, particularly when embedding table operations are offloaded to SSDs for embedding-dominated recommendation models. 
To bridge the performance gap, this paper proposes to use near data processing (NDP) by leveraging the existing compute capability of commodity SSDs.
Previous work has shown that NDP-based SSD systems can improve performance across a variety of different application spaces such as databases and graph analytics~\cite{ouyang2014sdf,schroeder2016flash}.
NDP solutions work particularly well when processing \textit{gather-reduce} operations over large quantities of input data using lightweight computations.
Embedding table operations follow this compute paradigm as well.
NDP can help reduce round-trip counts and latency overheads in PCIe communication as well as improve the SSD bandwidth utilization by co-locating compute with the Flash-based storage systems (more detail in Section~\ref{sec:design}).

In summary, \textit{the focus of this work is to demonstrate the viability of SSD-based storage for the  MLP-dominated recommendation models and to customize NDP-based SSD systems for neural recommendation in order to unlock the advantages of SSD storage capacity for the embedding-dominated models}.

%% file: design.tex
\section{\Name~Design}~\label{sec:design}
%
%
%
We present \Name, a {\it near-data processing} (NDP) solution for efficient embedding table operations on SSD memory. 
Compared to traditional SSD storage systems, \Name~increases bandwidth to Flash memories by utilizing internal SSD bandwidth rather than external PCIe,  greatly reducing unused data transmitted over PCIe by packing useful data together into returned logical blocks, and reduces command and control overheads in the host driver stack by reducing the number of I/O commands needed for the same amount of data. 
To maintain compatibility with exisitng NVMe protocol and drivers, \Name~is implemented within the FTL of the SSD, requiring no modifications to the hardware substrate and paving the way for datacenter scale deployment.
This section describes the overall \Name~design and implementation.
First we outline how embedding operations are mapped to the FTL in SSD systems; next, we detail how \Name~exploits temporal locality in embedding table operations to improve performance; and finally, we describe the feasibility of implementing \Name~in real systems. 



\subsection{Mapping Embedding Operations to the FTL}
\Name\ is designed to accelerate embedding table operations for recommendation inference. In most high-level machine learning frameworks, these embedding operations are implemented as specific custom operators. These operators can be implemented using a variety of backing hardware/software technologies, typically DRAM based data structures for conventional embedding operations. \Name\ implements embedding operations using SSD storage by moving computation into the SSD FTL, and on the host using custom NDP based drivers within the higher-level framework operator implementation.

Given the large storage requirements, embedding table operations (e.g., SparseLengthSum in Caffe2), span multiple pages within SSD systems. 
A standard FTL provides highly optimized software that supports individual page scheduling and maintenance; \Name~operates on top of request queues and data buffers designed for individual Flash page requests and operations. 
In order to support multi-page SparseLengthSum (SLS) operations, we add a scheduling layer -- with accompanying buffer space and request queues -- on top of the existing page scheduling layer. 
The proposed SLS scheduling layer feeds individual page requests from a set of in-flight SLS requests into the existing page-level scheduler to guarantee high throughput across SLS requests. 
The existing page-level scheduling proceeds as normal to ensure page operations maximize the available internal memory parallelism. 

\begin{figure*}[t]
    \centering
    \includegraphics[width=0.68\textwidth]{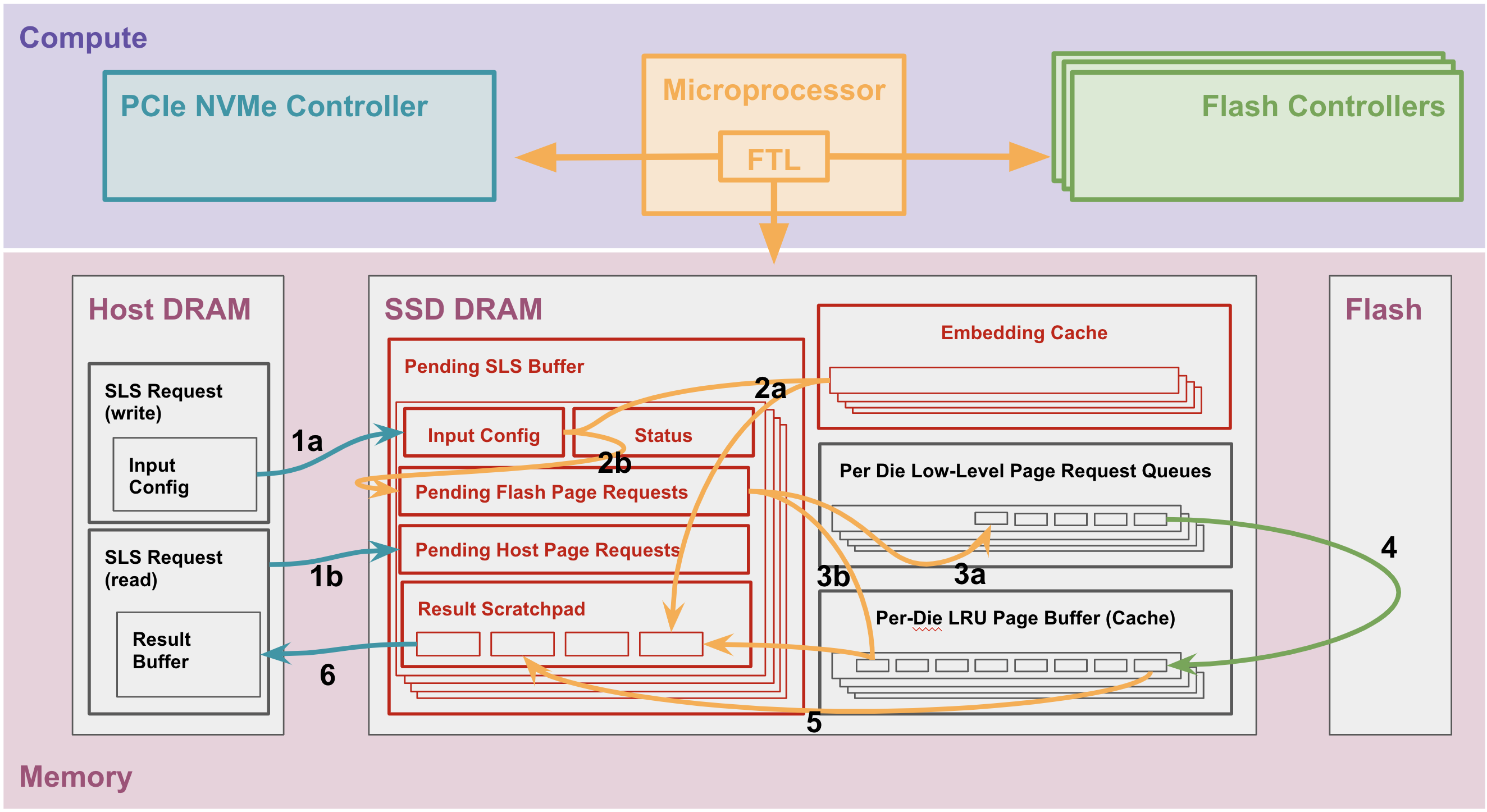}
    \caption{The lifetime of an SSD based SLS operator. The addition of an SLS request buffer and a specialized embedding cache support the multi-page operation. 
    }
    \label{fig:ndp_sls_flowchart}
\end{figure*}

Figure~\ref{fig:ndp_sls_flowchart} details the proposed \Name~ design, which augments SSD systems with NDP to improve internal Flash memory bandwidth and overall performance of embedding table operations.

\textbf{Data-structures} 
In particular, to support NDP SLS, \Name~adds two major system components: a pending-SLS-request buffer and a specialized embedding cache. These components are colored red in Figure~\ref{fig:ndp_sls_flowchart}.

Each SLS operation allocates an entry in the pending SLS request buffer. 
Each entry contains five major elements: (Input Config) buffer space to store SLS configuration data passed from the host, (Status) various data structures storing reformatted input configuration and counters to track completion status, (Pending Flash Page Requests) a queue of pending Flash page read requests to be submitted to the low-level page request queues, (Pending Host Page Requests) a queue of pending result logical block requests to be serviced to the host upon completion, and (Result Scratchpad) buffer space for those result pages. 

\textbf{Initiating embedding request} When the FTL receives an SLS request in the form of a write-like NVMe command, the FTL allocates an SLS request entry. 
The FTL then triggers the DMA of the configuration data from the host using the NVMe host controller (step 1a). Upon receipt of that configuration data, the FTL will need to process the data, initializing completion status counters and populating custom data structures containing the reformatted input data (populating element 2 - Status).
This processing step computes which flash Pages must be accessed and separates input embeddings by flash Page, such that the per-page processing computation can easily access its own input embeddings. During this scan of the input data, a fast path may also check for availability of input embeddings in an embedding cache (discussed later this section), avoiding flash Page read requests (step 2a), and otherwise placing those Flash page requests in the pending queue (step 2b). Upon completion of the configuration processing the request entry is marked as configured, and pending Flash page requests may be pulled and fed into the low-level page request queues (step 3a). If the page exists within the page cache already, the page may be directly processed (step 3b). When the FTL receives a SLS read-like NVMe command (asynchronous with steps 2-5), it searches for the associated SLS request entry and populates the pending host page request queue (step 1b).

\textbf{Issuing individual Flash requests} At a high level of the FTL scheduler polling loop, the scheduler will maintain a pointer to an entry in the SLS request buffer. Before processing low-level request queues, the scheduler will fill the queues from the current SLS entry's pending Flash page request queue. The scheduler will then perform an iteration of processing the low level page request queues, and increment the SLS request buffer pointer regardless of completion, such that requests are handled fairly in a round robin fashion.

\textbf{Returning individual Flash requests}  Upon completion of a Flash page read request which is associated with an SLS request (step 4), the extraction and reduction computation will be triggered for that page. The embeddings required for the request which reside in that page will be read from the page buffer entry and accumulated into the appropriate result embedding in the result buffer space for that SLS request (step 5). The reformatted input configuration allows the page processing function to quickly index which embeddings need to be processed and appropriately update the completion counters. 

\textbf{Returning embedding requests} Again at a high level of the FTL scheduler polling loop, the scheduler will check for completed host page requests within an SLS request. If completed pages are ready, and the NVMe host controller is available, the scheduler will trigger the controller to DMA the result pages back to the host (step 6). Upon completion of all result pages in a SLS request the SLS request entry will be deallocated. The NVMe host controller will automatically track completed pages and complete NVMe host commands.

\subsection{Exploiting temporal locality in embedding operations}

\textbf{Multi-threading and Pipelining} Aside from the base NDP implementation, there are a number of conventional optimizations that can be applied on top of the NDP Flash operation. Multi-threading and software pipelining can be used to overlap NDP SLS I/O operations with the rest of the neural network computation. For this we use a threadpool of SLS workers to fetch embeddings and feed post-SLS embeddings to neural network workers. We match our SLS worker count to the number of independent available I/O queues in our SSD driver stack. We then match our neural network workers to the available CPU resources.

DRAM caching is another technique which has been previously studied~\cite{eisenman2018bandana} in the context of recommendation inference serving. With our NDP implementation, there is the option for both host DRAM caching and SSD internal DRAM caching. 

\textbf{Host-side DRAM Caching} Because our NDP SLS operator returns accumulated result embeddings to the host, we cannot use our workload's existing NDP SLS requests to populate a host DRAM embedding cache. In order to still make use of available host DRAM, we implement a static partitioning technique utilizing input data profiling which can partition embedding tables such that frequently accessed embeddings are stored in host DRAM, while infrequently used embeddings are stored on the SSD. This solution is motivated by the characterization in Section~\ref{sec:characterize:locality}, showcasing the power law distribution of page access. Because there exist relatively few highly accessed embeddings, static partitioning becomes a viable solution. With this feature, our system requests the SSD embeddings using our NDP function, and post processes the returned partial sums to include embeddings contained in the DRAM cache on the host.

\textbf{SSD-side DRAM Caching}
For host DRAM caching, it is entirely feasible to use a large fully associative LRU software cache. However, for SSD internal DRAM caching, we must more carefully consider the implementation overheads of our software caching techniques. The FTL runs on a relatively simple CPU, with limited DRAM space. The code and libraries available are specifically designed for embedded systems, such that the code is compact and has low computation overhead, as well as having more consistent performance. The SSD FTL is designed without dynamic memory allocation and garbage collection. When implementing any level of associativity, the cost of maintaining LRU or pseudo LRU information on every access must be balanced against cache hit-rate gains. For the current evaluations we implement a direct-mapped SSD-side DRAM cache.


\subsection{Feasibility and Implementation: NDP SLS Interface}

Our custom interface maintains complete compatibility with the existing NVMe protocol, utilizing a single unused command bit to indicate embedding commands. Other than this bit, our interface simply uses the existing command structure of traditional read/write commands. Embedding processing parameters are passed to the SSD system with a special write-like command, which initiates embedding processing. A subsequent read-like command gathers the resulting pages. The parameters passed include embedding vector dimensions such as attribute size and vector length, the total number of input embeddings to be gathered, the total number of resulting embeddings to be returned, and a list of (input ID, result ID) pairs specifying the input embeddings and their accumulation destinations. Adding a restriction that this list be sorted by input ID enables more efficient processing on the SSD system, which contains a much less powerful CPU than the host system. The configuration-write command and result-read command are associated with each-other internally in the SSD by embedding a request ID into the starting logical block address (SLBA) of the requests. The SLBA is set as the starting address of the targeted embedding table added with the unique request ID. By assuming a minimum table size and alignment constraints, the two inputs can be separated within the SSD system using the modulus operator.

We also note that in addition to maintaining compatibility with existing NVMe protocol, by implementing support for embedding table operations purely through software within the SSD FTL, we ensure \Name~is fully compatible with existing commodity SSD hardware. 
This method of implementation relies on the lightweight nature of the required computation, such that the SSD microprocessor does not become overly delayed in its scheduling functions by performing the extra reduction computation. 

%% file: methodology.tex
\section{Methodology and Implementation}~\label{sec:methodology}
This section describes the methodology and experimental setup used to evaluate the proposed \Name~design.
Here we summarize the OpenSSD platform, Micron UNVMe, recommendation models, and input traces used.
Additional details can be found in the Appendix.

\textbf{OpenSSD}
In order to evaluate the efficacy of offloading the SLS operator onto the SSD FTL, we implement a fully functional NDP SLS operator in the open source Cosmos+ OpenSSD system~\cite{OpenSSD}. 
The development platform, Cosmos+ OpenSSD, has a 2TB capacity, fully functional NVMe Flash SSD, and a customizable Flash controller and FTL firmware.
In order to provide a feasible solution for datacenter scale deployment, we implement \Name within the FTL firmware; the interface is compatible with existing NVMe protocol, requiring no hardware changes.

\textbf{Micron UNVMe}
In addition to the NVMe compatible OpenSSD system, the \Name~interface is implemented within the Micron UNVMe driver library~\cite{unvme}. 
We modify the UNVMe driver stack to include two additional commands, built on top existing command structures for NVMe read/write commands and distinguished by setting an additional unused command bit, as described in Section~\ref{sec:design}.
The command interface enables flexible input data and command configurations, while maintaining compatibility with the existing NVMe host controller.
The UNVMe driver makes use of a low latency userspace library, which polls for the completion of NVMe read commands, and uses the maximum number of threads/command queues.


\textbf{Neural recommendation models}
To evaluate \Name, we use a diverse set of eight industry-representative recommendation models provided in DeepRecInfra~\cite{Gupta2020DeepRecSysAS}, implemented in Python using Caffe2~\cite{Caffe2}.
In order to evaluate the performance of end-to-end recommendation models on real systems, we integrate the SparseLengthsSum operations (embedding table operations in Caffe2) with the custom NDP solution. 
We offload embedding operations to \Name, we design a low-overhead Python-level interface using CTypes, which allows us to load the modified UNVMe as a shared library and call NVMe and NDP SLS I/O commands. 
In the future these operations could be ported into a custom Caffe2 operator function, and compiled along with the other Caffe2 C++ binaries. 

\textbf{Input traces and Performance Metrics}
In addition to the recommendation models themselves, we instrument the networks with synthetically generated input traces. 
We instrument the open-source synthetic trace generators from Facebook's open-scourced DLRM~\cite{naumov2019dlrm} with the locality analysis from industry-scale recommendation systems shown in Figure~\ref{fig:embedding_locality}.
The synthetic trace generator is instrumented with likelihood distributions for input embeddings across stack distances of previously requested embedding vectors. 
We generate exponential distributions based on a parameter value, $K$. 
Sweeping $K$ generates input traces with varying degrees of locality; for instance, setting $K$ equal to 0, 1, and 2 generates traces with 13\%, 54\%, and 72\% unique accesses respectively~\cite{gupta2019architectural, eisenman2018bandana}. 
Given the high cache miss rates and our locality analysis, we assume a single embedding vector per SSD page of 16KB. 
For the evaluation results, we assume embedding tables have 1 million vectors and host-side DRAM caches store up to 2K entries per embedding table.

Because our prototype limits us to single-model single-SSD systems, we do not focus our results on latency-bounded throughput, but rather direct request latencies, a critical metric for determining the performance viability of SSD based recommendation. We average latency results across many batches, ensuring steady-state behavior.





\textbf{Physical Compute Infrastructure}
All experiments are run on a Quad-core Intel Skylake desktop machine. Our machine uses G.SKILL TridentZ Series 64GB (4 x 16GB) 288-Pin DDR4 SDRAM DDR4 3200 (PC4 25600) Desktop Memory Model F4-3200C14Q-64GTZ DRAM. DRAM has nanosecond-scale latencies, and 10s of GB/s in throughput. Our prototype SSD system supports 10K IOPs per channel with 8 channels and a page size of 16KB, leading to maximum throughput with sequential read of just under 1.4GB/s. Newer SSD systems will have higher throughput. Single page access latencies are in the 10s to 100s of microseconds range.

%% file: evaluation.tex
\section{Evaluating \Name}~\label{sec:evaluation}
%
%
Here we present empirical results evaluating the performance of \Name.
Overall, the results demonstrate that \Name~provides up to 2$\times$ speedup over baseline SSD for recommendation inference.
This section first analyzes the fundamental tradeoffs of \Name using micro-benchmarks based on embedding table operations.
Following the micro-benchmark analysis, the section compares the performance of end-to-end recommendation models between baseline SSD systems and \Name.
Expanding upon this analysis, we present performance tradeoffs between baseline SSD systems and \Name~ using both host-side and SSD-side DRAM caching in order to exploit temporal locality in embedding table accesses. 
Finally, the section conducts a sensitivity study on the impact of individual recommendation network architectural parameters on \Name~performance.
The sensitivity analysis provides insight into \Name's performance on future recommendation models.

\begin{figure}[t]
    \centering
    \includegraphics[width=\linewidth]{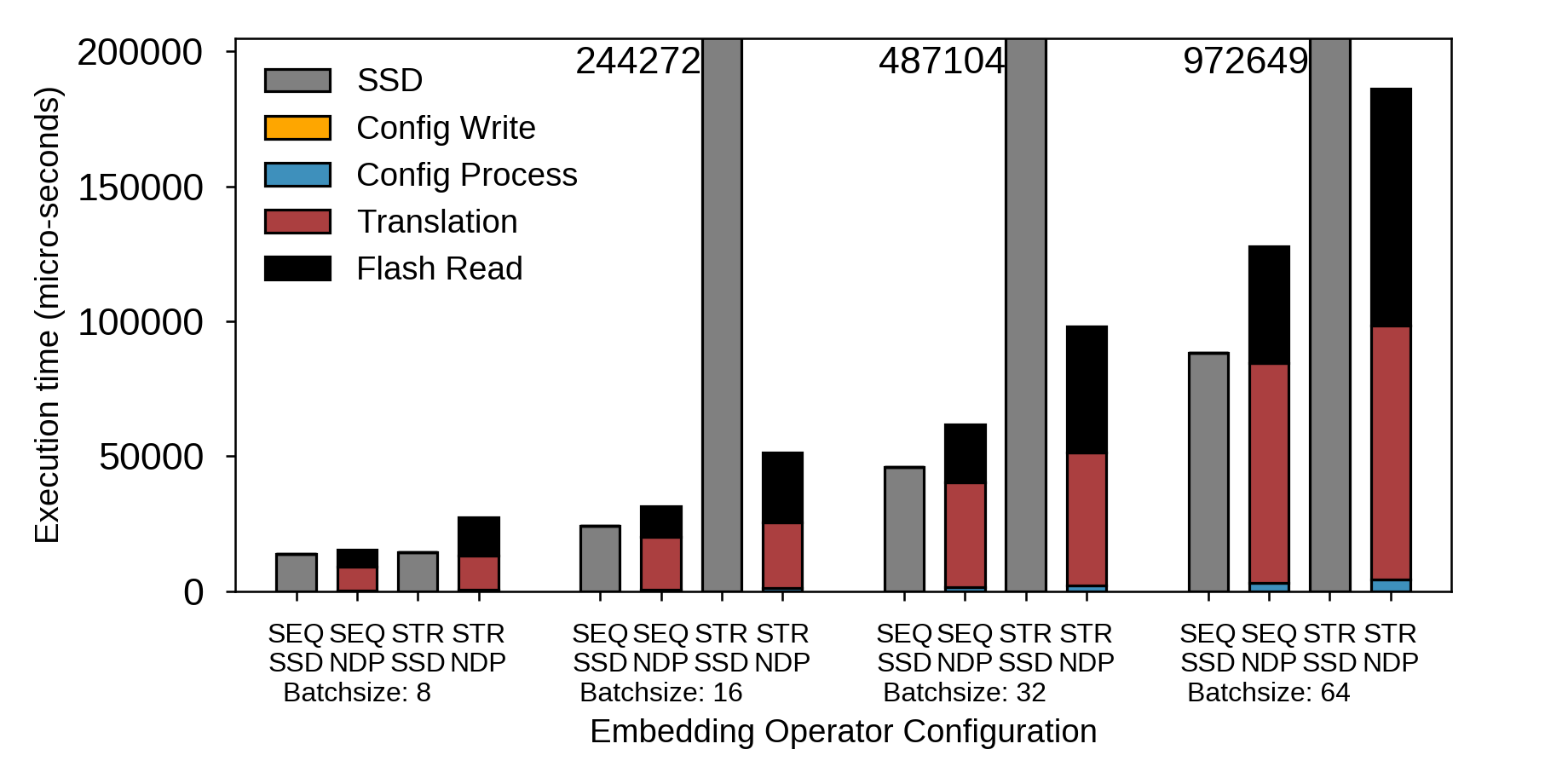}
    \caption{The standalone performance of the SLS embedding operator. Performance is shown for both sequential and strided access patterns, using both conventional SSD interfaces and NDP interfaces, on a variety of batch sizes.
    }
    \label{fig:ndp_sls_breakdown}
\end{figure}

\subsection{Accelerating embedding table operations}

Figure~\ref{fig:ndp_sls_breakdown} presents the performance of embedding table operations (i.e., SparseLengthsSum in Caffe2~\cite{Caffe2SLS}). 
For \Name, the execution time is categorized five components (i.e., Config Write, Config Process, Translation, and Flash Read) over a range of batch sizes.
\textit{Config Write} and \textit{Config Process} represent the time taken to transfer configuration data to the SSD and the time to process the configuration, respectively; after the transfers, internal data structures are populated and Flash page requests begin issuing. 
\textit{Translation} represents the time spent on processing returned Flash pages, extracting the necessary input embedding vectors, and accumulating the vectors into the corresponding result buffer space. 
\textit{Flash Read} indicates the time in which the FTL is managing and waiting on Flash memory operations.

It is difficult to compare the computational throughput of \textit{Translation} independently with the IO bandwidth of flash, as the computation is synchronously tightly-coupled with the IO scheduling mechanisms within the FTL. With hardware modification this computation could be decoupled and made parallel. However, we can indirectly observe the bottleneck by observing the dominating time spent in the FTL, whether it is translation computation or flash read operations.

Following the characterization from Section~\ref{sec:characterize:locality}, we study two distinct memory access patterns: SEQ and STR.
The \textit{Sequential}(SEQ) memory access pattern represents use cases where embedding table IDs are contiguous. 
This is unlikely to happen in datacenter-scale recommendation inference applications, as shown in Figure~\ref{fig:embedding_locality}, but represents use cases with extremely high page locality.
The \textit{Random}(STR) memory access patterns are generated with strided embedding table lookup IDs and representative of access patterns where each vector accessed is located on a unique Flash page.
Given the large diversity in recommendation use cases, as evidenced by the variety of recommendation model architectures~\cite{Gupta2020DeepRecSysAS}, the two memory access patterns allow us to study the performance characteristics of \Name~across a wide variety use cases. 
Furthermore, while current recommendation use cases exhibit sparse access patterns, future optimizations in profiling and restructuring embedding tables may increase the locality.



\textbf{Performance with low locality embedding accesses}
Under the \textit{Random} memory lookup access pattern, \Name~achieves up to a 4$\times$ performance improvement over baseline SSD. 
This performance improvement comes from the increased memory level parallelism.
\Name~increases memory level parallelism by concurrently executing Flash requests for each embedding operation, increasing utilization of the internal SSD bandwidth. 
As shown in Figure~\ref{fig:ndp_sls_breakdown}, roughly half the time in the \Name's FTL is spent on \textit{Translation}. 
Given the limited hardware capability of the 1GHz, dual core ARM A9 processors of the Cosmos+OpenSSD system\cite{OpenSSD}, we expect that with faster SSD microprocessors or custom logic, the \textit{Translation} time could be significantly reduced.

\textbf{Performance with high locality embedding accesses}
Compared to the baseline SSD system using conventional NVMe requests, \textit{Sequential} access patterns with high spatial locality result in poor NDP performance. 
Compared to random or low locality access patterns, sequential or high locality embedding accesses access fewer Flash pages but require commensurate compute resources to aggregate embedding vectors. 
In the baseline system, the SSD page cache will hold pages while embedding vectors requests are sequentially streamed through the system and accumulated by the CPU.
While, \Name~also acccess fewer Flash pages, the embedding vectors are aggregated using the dual-core ARM A9 processor on the Cosmos+OpenSSD system; this accounts for over half the execution time (Translation) as shown in Figure~\ref{fig:ndp_sls_breakdown}.
With sequential accesses, the benefits of aggregating faster server class, host Intel CPU outweighs the lack of overhead of multiple NVMe commands.
We anticipate more sophisticated processors on the NDP system would close eliminate the slight performance degradation using \Name.

\begin{figure}[t]
    \centering
    \includegraphics[width=0.5\textwidth]{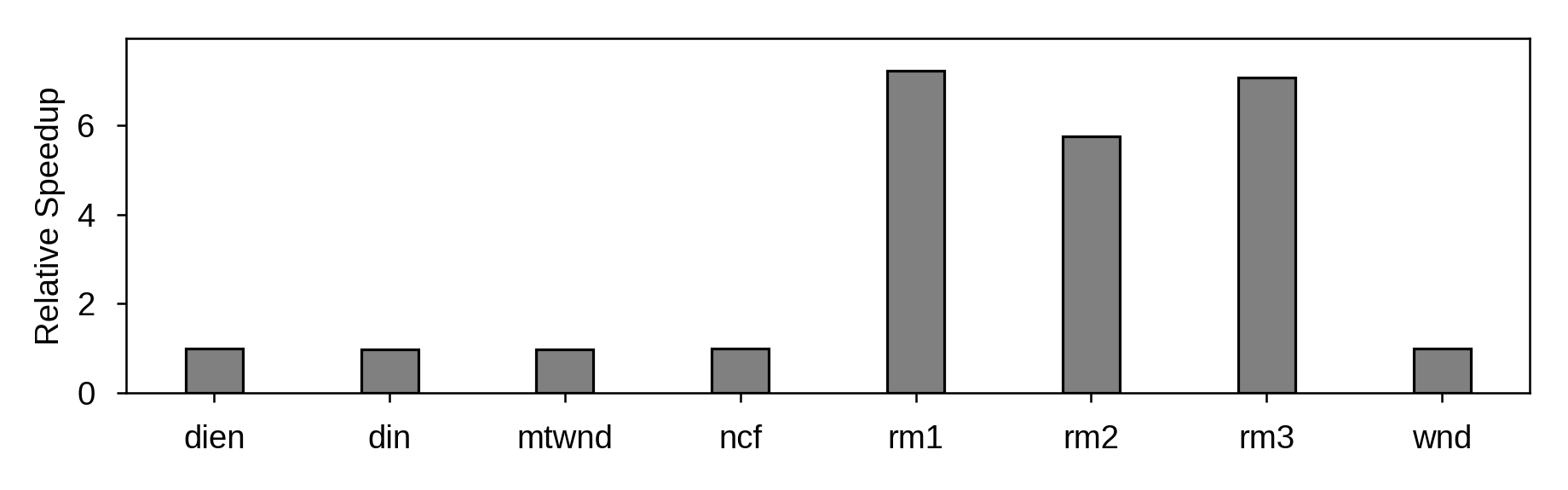}
    \caption{
    NDP alone provides up to 7$\times$ performance improvement for some full models given a simple naive configuration. 
    }
    \label{fig:full_model_naive}
\end{figure}

\begin{figure*}[ht]
    \centering
    \includegraphics[width=\textwidth]{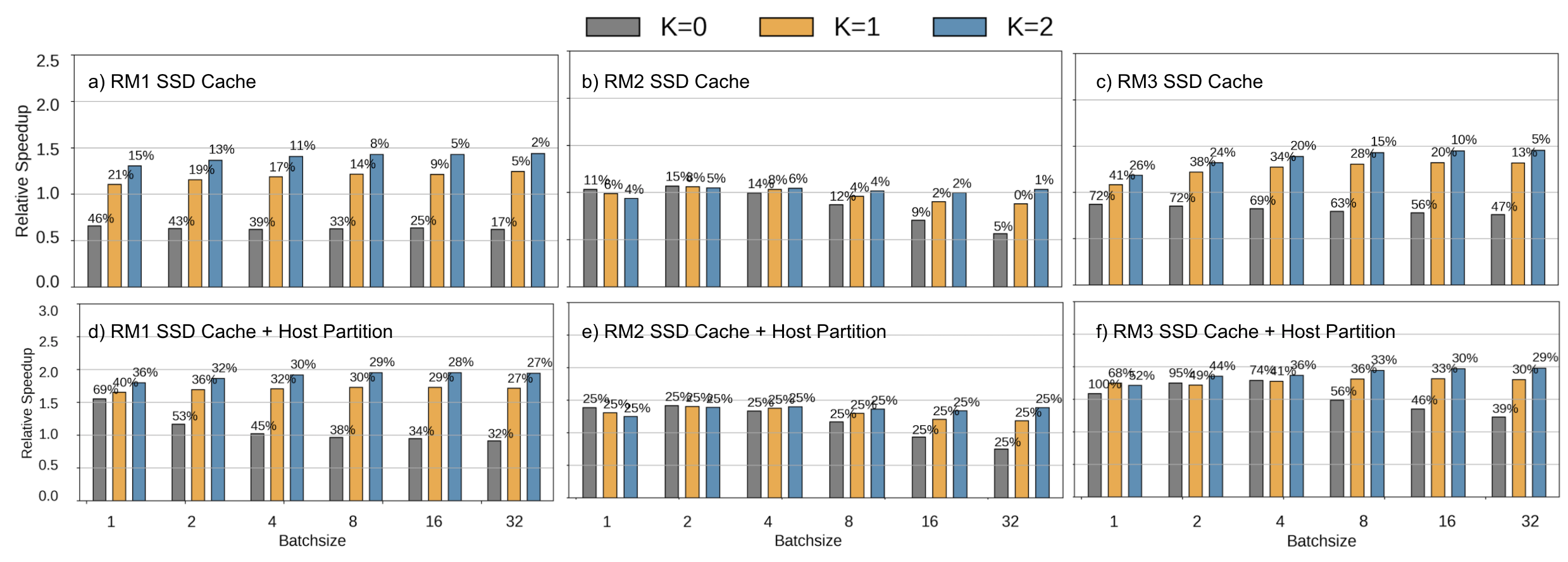}
    \caption{
    Relative full model performance improvement including caching techniques. The percentages above each bar represent the hitrate of either the SSD cache (a-c) or the host partition (d-f) for RecSSD.
    The baseline LRU cache hitrates always follow the inverse of the locality distribution, with 84\%, 44\%, and 28\% hits in the cache corresponding to $K$ equal to 0, 1, and 2, respectively.
    }
    \label{fig:full_model_performance}
\end{figure*}

\subsection{Accelerating end-to-end recommendation models}
In addition to individual embedding table operations, here we evaluate \Name~across a selection of industry-representative recommendation models.
To start, we showcase the raw potential of NDP, by presenting the simplest naive experimental configuration. Figure~\ref{fig:full_model_naive} presents the relative speedup of \Name~over a conventional SSD baseline, without operator pipelining and caching techniques, and using randomly generated input indices. 
We observe that many models exist where NDP provides no observable benefits, and for models where performance is limited by embedding operations and SSD latencies, NDP can provide substantial assistance with up to 7$\times$ speedup. 
The maximum speedup across models shown here is higher than that of the individual embedding operations (Figure~\ref{fig:ndp_sls_breakdown}) due to differences in underlying model parameters such as feature size and indices per lookup as discussed in Section~\ref{sec:params}.

\subsection{Exploiting Locality in end to end models}
In addition to the end-to-end model results, we evaluate the performance of \Name~with operator pipelining and caching.
These optimization techniques, as presented in Section~\ref{sec:design}, are applied on top of \Name~and conventional SSD systems. 
Figure~\ref{fig:full_model_performance}(a-c) presents relative speedup results for \Name~with just SSD-side caching and the conventional SSD baseline with host-side caching. 
\Name~ utilizes a large, but direct mapped, cache within the SSD DRAM while the baseline utilizes a fully associative LRU cache within host DRAM.
Batchsizes are swept between 1 and 32, along with the three input trace locality conditions $K=0,1,2$. 
Hit rates for \Name's SSD DRAM cache are labeled above each speedup bar. 
The baseline LRU cache hit rates follow the inverse of the locality distribution, with 84\%, 44\%, and 28\% hits in the cache corresponding to $K$ equal to 0, 1, and 2 respectively. 
Note, the LRU cache hit rates span the diverse set of embedding access patterns from the initial characterization of production-scale recommendation models shown in Figure~\ref{fig:embedding_locality}.  

With high locality (i.e., low $K$), conventional SSD systems achieve higher performance than \Name.
On the other hand, with low locality \Name~outperforms the conventional baseline. 
This is because the direct mapped caching hit rate cannot match that of the more complex fully associative LRU cache on the host system, exemplified in the high batch size runs for RM1. 
Furthermore, RM2 has lower SSD cache hit rates compared to RM1/3, a result of the larger number of embedding lookups required per request and temporal locality being across requests not lookups. 
Even so, without host DRAM caching, \Name~ outperforms the baseline by up to 1.5X for lower locality traces, where many SSD pages must be accessed and the benefits of increased internal bandwidth shine. 

Figure~\ref{fig:full_model_performance}(d-f) presents relative speedup results for \Name~using static table partitioning as well as SSD caching. 
With static table partitioning, we make use of available host DRAM by statically placing the most frequently used embeddings within the host DRAM cache as detailed in Section~\ref{sec:design}. 
The hitrates labeled above each bar represent the hit rates of \Name~ in the statically partitioned host DRAM cache, not the SSD cache.

Following the conventional SSD baseline, static partitioning helps in leveraging the available host DRAM memory. 
For high temporal locality however it cannot match the hit rate of the fully associative LRU cache. 
With higher batch sizes as well as higher indicies per request (seen in RM2), the hit rate asymptotically approaches 25\%, the size of the static partition relative to the used ID space.
Overall, Figure\ref{fig:full_model_performance} shows that with static partitioning, \Name~achieves a 2$\times$ performance improvement over the conventional SSD baseline. 
This occurs when the baseline host LRU cache has a relatively low hit rate such that many SSD pages must be accessed, while \Name~is able to achieve comparable host DRAM hitrates with static partitioning.

In general, the above results show that the advantages of \Name\ shine when pages must be pulled from the SSD, and when the host level caching strategies available for \Name\ (static partitioning) are of comparable effectiveness to the baseline LRU software cache. Although \Name\ shows diminishing returns with improved caching and locality, we note that because \Name\ is fully compatible with the existing NVMe interface, it can be employed in tandem with conventional strategies and switched based on the embedding table locality properties.


\subsection{Sensitivity study: Impact of model architecture}~\label{sec:params}
In this section we more closely examine the impact of model parameters differentiating the performance of our benchmark models. Table~\ref{tab:bench_parameters} details the parameter space of RM1/2/3. We specifically note that absolute table size does not impact our results. Growing table sizes do provide motivation to move from capacity constrained DRAM to flash SSDs, however embedding lookup performance is dependant on access patterns, not absolute table size.

\begin{table}[ht]
    \centering
    \caption{Differentiating benchmark parameters.}
    \label{tab:bench_parameters}
    \begin{tabular}{c | c c c } 
    \toprule
    Benchmark & Feature Size & Indices & Table Count \\ [0.5ex] 
    \midrule
    RM1 & 32 & 80 & 8 \\ 
    
    RM2 & 64 & 120 & 32 \\
    
    RM3 & 32 & 20 & 10 \\
    \bottomrule
    \end{tabular}
\end{table}

\begin{figure}[t]
    \subfloat[Feature Size and Quantization]{
        \begin{minipage}[c][1\width]{0.22\textwidth}
            \centering
            \includegraphics[width=\textwidth]{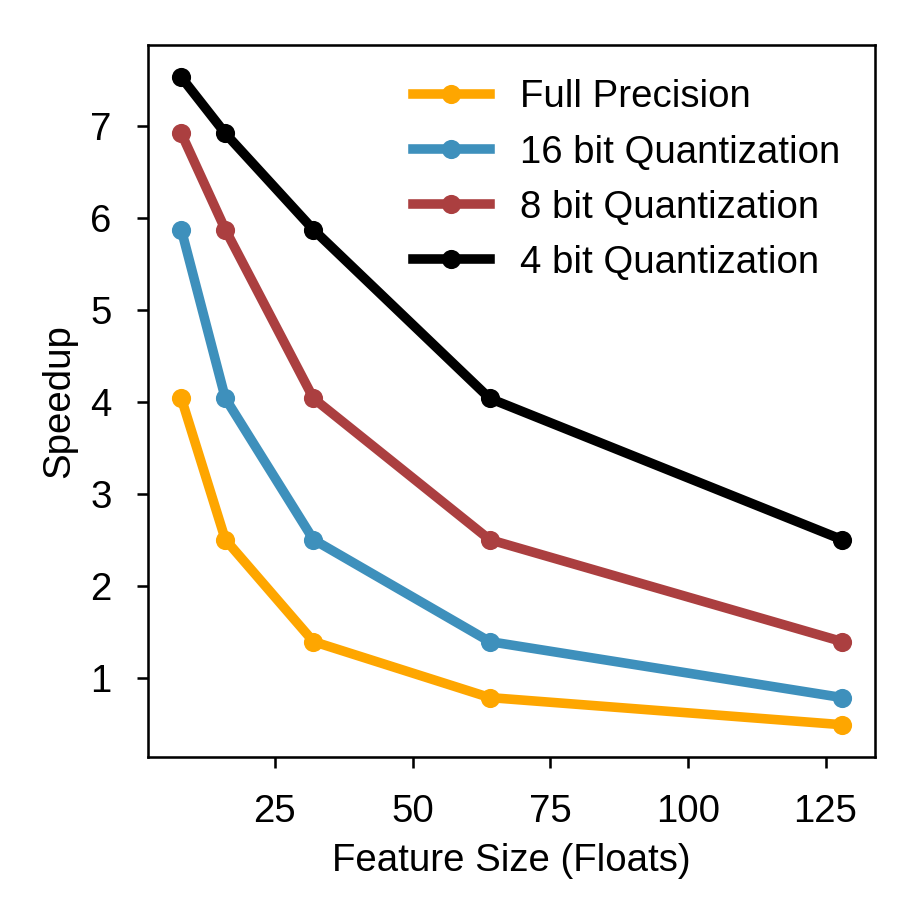}
            \label{fig:featuresize}
        \end{minipage}
    }
    \hfill
    \subfloat[Indices and Table Count]{
        \begin{minipage}[c][1\width]{0.22\textwidth}
            \centering
            \includegraphics[width=\textwidth]{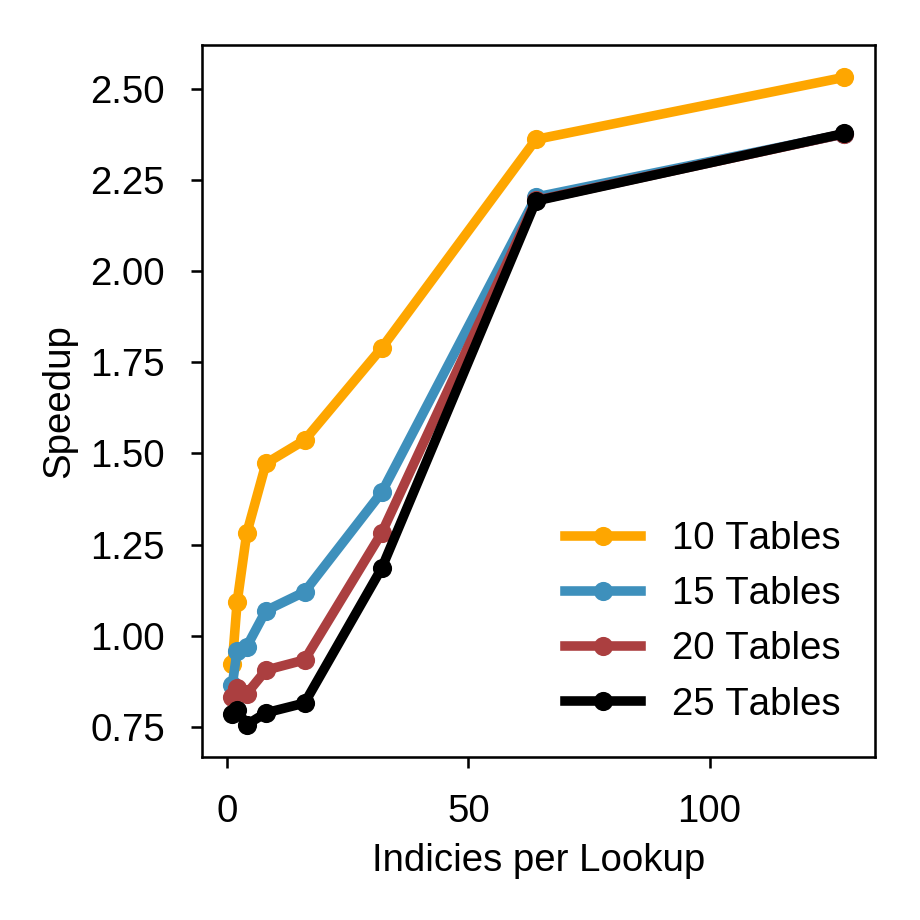}
            \label{fig:indicies}
        \end{minipage}
    }
    \caption{Examining the impact of model parameters on full model executions. 
    } 
    \label{fig:full_model_latency}
\end{figure}

In Figure~\ref{fig:featuresize} we see how feature size and quantization, which affect the size of embedding vectors relative to the page size, show decreasing relative performance as this ratio grows. This is because the baseline is able to make more efficient use of block accesses as the lowest unit of memory access approaches the size of a memory block, while \Name~must perform more computation on the SSD microprocessor per page accessed from Flash. In Figure~\ref{fig:indicies} we see that although increasing table count diminishes performance, this quickly becomes outscaled by increases in performance from the increased indices per lookup. The performance loss from increasing table count is due to the implementation of our NDP interface. Because a single NDP call handles a single table, the amortization of command overheads is on a per table basis. On the other hand, increasing the number of indices per lookup increases the amortization of this control overhead as well as the value of accumulating these embeddings on the SSD, where only one vector is sent to the host for all the indices accumulated in a single lookup. 




%% file: related.tex
\section{Related Work}~\label{sec:related}
%
\textbf{SSD systems}
Recent advances in flash memory technology have made it an increasingly compelling storage system for at-scale deployment.
Compared to disk based solutions, SSDs offer 2.6$\times$ and 3.2$\times$ bandwidth per Watt and bandwidth per dollar respectively~\cite{andersen2010rethinking}.
Furthermore, given the high density and energy efficiency, SSDs are being used as datacenter DRAM-replacements as well~\cite{andersen2010rethinking}.
In fact, prior work from Google and Baidu highlight how modern SSD systems are being used for web-scale applications in datacenters~\cite{ouyang2014sdf,schroeder2016flash}. 
Furthermore, given recent advances in Intel's Optane technology, balancing DRAM-like latency and byte-addressability with SSD-like density, we anticipate the type of applications that leverage SSD based storage systems to widen~\cite{izraelevitz2019basic}.
In fact, training platforms for terabyte scale personalized recommendation models rely heavily on SSD storage capabilities for efficient and scalable execution~\cite{Zhao2020DistributedHG}. 

In order to enable highly efficient SSD execution, modern storage solutions rely on programmable memory systems~\cite{do2019programmable}. Leveraging this compute capability, there has been much work on both software and hardware solutions for Near Data Processing in SSDs for other datacenter applications~\cite{DoSmartSSDs2013,Willow,wang2016ssd,ouyang2014sdf,Biscuit,KAML,LSMSSD,ActiveFlash}.
Previous works which target more general SSD NDP solutions have relied on hardware modifications, complex programming frameworks, and heavily modified driver subsystems to support the performance requirements of more complex and general computational tasks. Our system trades-off this generality for simplicity and application specific performance and cost efficiency.


\textbf{Accelerating recommendation inference}
Given the ubiquity of AI and machine learning workloads, there have been many proposals for accelerating deep learning workloads. In particular, recent work illustrates that recommendation workloads dominate the AI capacity in datacenters~\cite{gupta2019architectural}.
As a result, recent work proposes accelerating neural recommendation.
For example, the authors in~\cite{kwon2019tensordimm,Hyun2019NeuMMUAS} propose a customized memory management unit for AI accelerators (i.e., GPUs) in order to accelerate address translation operations across multi-node hardware platforms.
Given the low-compute intensity of embedding table operations, recent work also explores the role of near memory processing for Facebook's recommendation models~\cite{Ke2019RecNMPAP}.
Similarly, researchers have proposed the application of flash memory systems to store large embedding tables found in Facebook's recommendation models~\cite{eisenman2018bandana}, exploring advanced caching techniques to alleviate challenges with large flash page sizes. These techniques can be used in tandem with \Name.
In this paper, we explore the role combining near data processing and NAND flash memory systems for at-scale recommendation in order to reduce overall infrastructure cost.
Furthermore, we provide a real system evaluation across a wide collection of recommendation workloads~\cite{Gupta2020DeepRecSysAS}.

%% file: conclusion.tex
\section{Conclusion}
In this paper we propose, \Name, a near data processing solution customized for neural recommendation inference.
By offloading computations for key embedding table operations, \Name~reduces round-trip time for data communication and improves internal SSD bandwidth utilization.
Furthermore, with intelligent host-side and SSD-side caching, \Name~enables high performance embedding table operations. 
We demonstrate the feasibility of \Name~by implementing it in a real-system using server-class CPUs and an OpenSSD compatible system with Micron UNVMe's driver library. 
\Name~reduces end-to-end neural recommendation inference latency by 4$\times$ compared to off-the-shelf SSD systems and comparable performance to DRAM-based memories.
As a result, \Name~enables highly efficient and scalable datacenter neural recommendation inference. 

%% file: artifact.tex
\section{Artifact Appendix}

\subsection{Abstract}

\Name\ is composed of a number of open sourced artifacts. First, we implement a fully-functional NDP SLS operator in the open source Cosmos+ OpenSSD system~\cite{OpenSSD}, provided in the \emph{RecSSD-OpenSSDFirmware} repository\cite{recssd-openssdfirmware}. 
To maintain compatibility with the NVMe protocols, the \Name\ interface is implemented within Micron's UNVMe driver library~\cite{unvme}, provided in the \emph{RecSSD-~UNVMeDriver} repository\cite{recssd-unvmedriver}. To evaluate \Name, we use a diverse set of eight industry-representative recommendation models provided in DeepRecInfra~\cite{Gupta2020DeepRecSysAS}, implemented in Python using Caffe2~\cite{Caffe2} and provided in the \emph{RecSSD-RecInfra} repository\cite{recssd-recinfra}.
In addition to the models themselves, we instrument the open-source synthetic trace generators from Facebook's open-sourced DLRM~\cite{naumov2019dlrm} with our locality analysis from production-scale recommendation systems, also included in the RecSSD-RecInfra repository.

\subsection{Artifact check-list (meta-information)}


{\small
\begin{itemize}
  \item {\bf Compilation: } GCC, Python3, PyTorch, Caffe2, Xilinx SDK 2014.4
  \item {\bf Model: } DeepRecInfra
  \item {\bf Run-time environment: } Ubuntu 14.04
  \item {\bf Hardware: } Cosmos+ OpenSSD, two Linux Desktop machines, remote PDU
  \item {\bf How much time is needed to prepare workflow (approximately)?: } 4-8 hours, once hardware is acquired
  \item {\bf How much time is needed to complete experiments (approximately)?: } 10+ hours
  \item {\bf Publicly available?: } Software will be open-sourced and publicly available. Required hardware platform is potentially still purchasable through original developers.
  \item {\bf Code licenses (if publicly available)?: } GNU GPL
\end{itemize}}

\subsection{Description}

\subsubsection{How to access}

\Name\ is provided through a number of publically available GitHub repositories~\cite{recssd-openssdfirmware,recssd-recinfra,recssd-unvmedriver}, as well as a publicly available archive on Zenodo, DOI: 10.5281/zenodo.4321943.




\subsubsection{Hardware dependencies}

Cosmos+ OpenSSD system~\cite{OpenSSD}, two Linux Desktop class machines, and a remote PDU for a fully remote workflow.

\subsubsection{Software dependencies}

Xilinx SDK 2014.4 for programming the OpenSSD. The Cosmos+ OpenSSD FTL firmware and controller Bitstream. Python3, PyTorch, and Caffe2 for running recommendation models.


\subsubsection{Models}

Uses recommendation model benchmarks from DeepRecInfra~\cite{Gupta2020DeepRecSysAS}, and trace generation from Facebook's open-sourced DLRM~\cite{naumov2019dlrm}.

\subsection{Installation}

To set up the SSDDev machine, start by downloading the Cosmos+ OpenSSD software available on their GitHub\cite{openssd-github}. You will need to install Xilinx SDK 2014.4, and follow the instructions in their tutorial\cite{openssd-tutorial} to set up a project for the OpenSSD board. For RecSSD, we use the prebuilt bitstream and associated firmware. After setting up the project, replace the ./GreedyFTL/src/ directory with the code from the RecSSD-OpenSSDFirmware GitHub repository. The OpenSSD tutorial contains detailed instructions on running the firmware, and the physical setup of the hardware.

To set up the SSDHost machine, download and make the RecSSD-UNVMeFirmware repository. This repository provides a user level driver library to connect the RecSSD-RecInfra recommendation models to the OpenSSD device. Once the SSDHost has been booted with the OpenSSD running, use lspci to detect the PCIe device identifier of the board, and use unvme-setup bind PCIEID to attach the driver to the specific device. Make note of ./test/unvme/libflashrec.so, which must be later copied into RecSSD-RecInfra, such that the Python3 runtime can load and run the necessary driver functions to make use of our implemented NDP techniques.

Next, download the RecSSD-RecInfra repository. Copy the libflashrec.so file into ./models/libflashrec.so. Make sure to download and install Python3 and PyTorch\cite{pytorch}.

\subsection{Experiment workflow}

Detailed walk-throughs of the technical steps required are documented within the provided individual repositories and from the OpenSSD tutorial\cite{openssd-tutorial}. At a high level the expected workflow is as follows.

\begin{enumerate}
    \item With the SSDHost machine powered off, use the Xilinx SDK on the SSDDev machine to launch the FTL firmware on the OpenSSD.
    \item Power on and boot the SSDHost machine. Connect the UNVMe driver library to the device through unvme-setup bind.
    \item Run /models/input/create\_dist.sh within RecSSD-RecInfra to generate the desired synthetic locality patterns for input traces.
    \item Run the python based experimental sweeps scripts within RecSSD-RecInfra /models/ to run various recommendation models using either baseline SSD interfaces or our NDP interfaces.
\end{enumerate}

\subsection{Evaluation and expected results}

Most of our results are reported as inference latency, output from scripts run on the SSDHost machine. We compare relative latency results across a large number of batches in order to guarantee regular steady state behavior. Figure~\ref{fig:full_model_performance} presents expected results for the important RM1, RM2, and RM3 models, while Figure~\ref{fig:full_model_latency} presents results for an RM3-like model while tuning specific model parameters.

Figure~\ref{fig:ndp_sls_breakdown} reports breakdowns in time spent within the FTL for NDP requests using microbenchmarks within the RecSSD-UNVMeDriver repository. To reproduce these results, run ./test/unvme/unvme\_embed\_test. Unlike model latency results, these measurements are performed within the FTL and directly reported through output to the SDK, therefore they must be recorded from the SDK running on the SSDDev machine.

Figures~\ref{fig:embedding_example0} and~\ref{fig:embedding_locality} use proprietary industry data and are not reproducible using our open-sourced infrastructure.




